# Detection of Chinese Stock Market Bubbles with LPPLS Confidence Indicator

Min Shu[1, 2, *], Wei Zhu[1, 2]

[1] Department of Applied Mathematics & Statistics, Stony Brook University, Stony Brook, NY, USA
[2] Center of Excellence in Wireless & Information Technology, Stony Brook University, Stony Brook, NY, USA

## Abstract

We present an advance bubble detection methodology based on the Log Periodic Power Law Singularity (LPPLS) confidence indicator for the early causal identification of positive and negative bubbles in the Chinese stock market using the daily data on the Shanghai Shenzhen CSI 300 stock market index from January 2002 through April 2018. We account for the damping condition of LPPLS model in the search space and implement the stricter filter conditions for the qualification of the valid LPPLS fits by taking account of the maximum relative error, performing the Lomb log-periodic test of the detrended residual, and unit-root tests of the logarithmic residual based on both the Phillips-Perron test and Dickey-Fuller test to improve the performance of LPPLS confidence indicator. Our analysis shows that the LPPLS detection strategy diagnoses the positive bubbles and negative bubbles corresponding to well-known historical events, implying the detection strategy based on the LPPLS confidence indicator has an outstanding performance to identify the bubbles in advance. We find that the probability density distribution of the estimated beginning time of bubbles appears to be skewed and the mass of the distribution is concentrated on the area where the price starts to have an obvious super-exponentially growth. This study is the first work in the literature that identifies the existence of bubbles in the Chinese stock market using the daily data of CSI 300 index with the advance bubble detection methodology of LPPLS confidence indicator. We have shown that it is possible to detect the potential positive and negative bubbles and crashes ahead of time, which in turn limits the bubble sizes and eventually minimizes the damages from the bubble crash.

Keywords: Financial bubble; Market crash; Log-periodic power law singularity (LPPLS); Chinese stock market; Confidence indicator; Bubble indicator

---
*Corresponding author at: Department of Applied Mathematics & Statistics, Physics A149, Stony Brook University, Stony Brook, NY 11794, USA.
*E-mail address:* min.shu@stonybrook.edu (M. Shu), wei.zhu@stonybrook.edu (W. Zhu)

## 1. Introduction

In the modern society, financial bubbles and crashes are not rare phenomena and have great impact on the lives and livelihoods of most people all over the world. Approximately 100 financial crises worldwide have been observed in the past 30 years [1]. It is vital to identify bubbles in advance, limit their sizes, and eventually minimize the damage from the bubble crash. The causes of bubbles have been widely investigated and recent theories indicate bubbles of stock market are generated because of (1) heterogeneous beliefs of investors together with short-time constraints, (2) positive feedback trading by noise traders, and (3) synchronization failures among rational traders [2].

In order to detect the presence of a bubble effectively, the Log Periodic Power Law Singularity (LPPLS) model [3-5] has been developed at the interface of financial economics, behavioral finance and statistical physics. In the LPPLS model based on the theory of rational expectation, the bubbles are believed to be characterized by faster-than-exponential (or super-exponential) growth of price leading to unsustainable growth ending with a finite crash-time $t_c$. The super-exponential growth of price of a bubble results from positive feedback mechanism in the valuation of assets created by imitation and herding behavior of noise traders and of boundedly rational agent results in price processes that exhibit a finite-time singularity at some future time [2]. Because of the tension and competition between the value investors and the noise traders, the market price of an asset is deviated around the faster-than-exponential growth in the form of oscillations that are periodic in the logarithm of time to $t_c$. Based on analyzing the price time series of an asset, the LPPLS model provides a flexible framework to detect financial bubbles. Over the past decade, the LPPLS model has been widely used to detect bubble and crashes in various markets, such as the real estate market in Las Vegas [6], the 2000-2003 real estate bubble in the UK [7], the USA real estate bubble [8], the 2006-2008 oil bubble [9], the Chinese stock market bubbles in 2005–2007 and 2008–2009 [10], and the Shanghai 2015 stock market bubble [11].

In recent year, there is a growing research on using the LPPLS model to detect bubbles. Yan, Woodard and Sornette [12] adapted the LPPLS formula to model the negative bubbles, so that the market rebounds can be detected through pattern recognition. Brée, Challet and Peirano [13] found that the LPPLS functions are intrinsically hard to fit to time series by accounting for the sloppiness. Sornette, Woodard, Yan and Zhou [14] discussed the theoretical status and common calibration issues concerning the LPPLS model. Filimonov and Sornette [15] transformed the formulation of the LPPLS formula to reduce the number of nonlinear parameters in the function from four to three, which reduces the complexity and improves the stability of the calibration. Geraskin and Fantazzini [16] presented a detailed guide for modelling and identifying financial bubbles using the LPPLS model. Lin, Ren and Sornette [17] proposed a self-consistent model for explosive financial bubbles that combines a mean-reverting volatility process with a stochastic conditional return. Sornette, Demos, Zhang, Cauwels, Filimonov and Zhang [11] evaluated the performance of the real-time prediction of the bubble crash in 2015 Shanghai stock market by constructing the LPPLS Confidence indicator and the LPPLS Trust indicator, and conducted the relevant post-mortem analysis on the effectiveness of LPPLS methodology. Zhang, Zhang and Sornette [18] adopted the quantile regression for LPPLS calibration and used a multi-scale analysis to combine the many quantile regressions. The LPPLS confidence and trust indicators were also implemented to enrich the diagnostic of bubbles. Li [19] investigated the critical times of three historical Chinese stock market bubbles confirming that the LPPLS performs well in predicting the bubble crashes and the forecast gap is an alternative way for the market conversion warning. Demos and Sornette [20] carried out systematic tests of the precision and reliability of determining the beginning and end time of a bubble, and found that the beginning of bubbles is much better constrained than their end. Filimonov, Demos and Sornette [21] applied the modified profile likelihood inference method to calibrate the LPPLS model for financial bubbles and obtained the interval estimation for the critical time. Demirer, Demos, Gupta and Sornette [22] applied the LPPLS confidence multi-scale indicators to evaluate the predictive power of market-based indicators and identified that short selling and liquidity are two important factors contributing to the bubble indicators.

The financial system in China have evolved from Mao's single-bank system to Deng's four-bank system and till now, is still dominated by its state-owned bank sector. China's stock markets opened in 1990, mainly as a platform for privatization of state-owned enterprises and the selected firms in the list strictly controlled by the government. Until 2005, only one-third of the equity shares were tradable, and the total market capitalization was below $1 trillion until 2006 [23]. On the strength of a series of developments over the last decade, the Chinese economy has a stellar growth and Chinese GDP has more than tripled to over $11 trillion in 2016. The capitalization of the Chinese stock market has grown more than five-fold to over $7 trillion by May 2017, and the Chinese stock market rose to the world's second largest, attracting attention from mainstream research in financial economics.

With the rapid growth of the Chinese economy, China's stock markets have experienced a roller coaster dynamics, with three large bubbles bursting from May 2005 to October 2007, from November 2008 to August 2009, and from mid-2014 to June 2015 [11]. In mainland China, the organized stock market is composed of two stock exchanges: the Shanghai stock exchange (SHSE) and the Shenzhen stock exchange (SZSE). One of the most important indices for A-shares is the Shanghai Shenzhen CSI 300 index (CSI 300), which is a capitalization-weighted stock market index representing the performance of the top 300 stocks traded in the Shanghai and Shenzhen stock exchanges. The CSI 300 index has been calculated since April 8, 2005. The evolution of the price trajectories of the CSI 300 index is shown in Figure 1. In the Chinese stock bubble of 2007, the CSI 300 index soared 573.2% from 873 on December 1, 2015 to 5,877.2 on October 16, 2007, and then the CSI 300 index suffered a more than 70% drop from the historical high during the period from October 2007 to October 2008. The 2015 Chinese Stock Market bubble crashed on June 12, 2015. The CSI 300 index has lost more than 42% from the peak on June 12, 2015 to the bottom on August 26, 2015.

In this study, we adopt the LPPLS methodology to detect the positive and negative bubbles in the Chinese stock market using the daily data on the CSI 300 stock market index from January 2002 through April 2018. This study is the first work in the literature that identifies the existence of bubbles in the Chinese stock market using the daily data of CSI 300 index with the advance bubble detection methodology of LPPLS confidence indicator. To improve the performance of LPPLS confidence indicator, the damping condition of LPPLS model is included in the search space and the stricter filter conditions for the qualification of the valid LPPLS fits are applied in this study. This study also presents the additional results on the two "well-known" Chinese stock market bubbles occurring on 2007 and 2015 respectively to demonstrate the LPPLS methodology for detecting the bubbles and their terminations.

The paper is organized as follows. Section 2 presents the technical descriptions of all the methods used in this study, including the LPPLS model, LPPLS calibration, and LPPLS confidence indicator. Section 3 contains the empirical analysis of the LPPLS confidence indicator application to the Chinese Stock Market; while Section 4 concludes this paper.

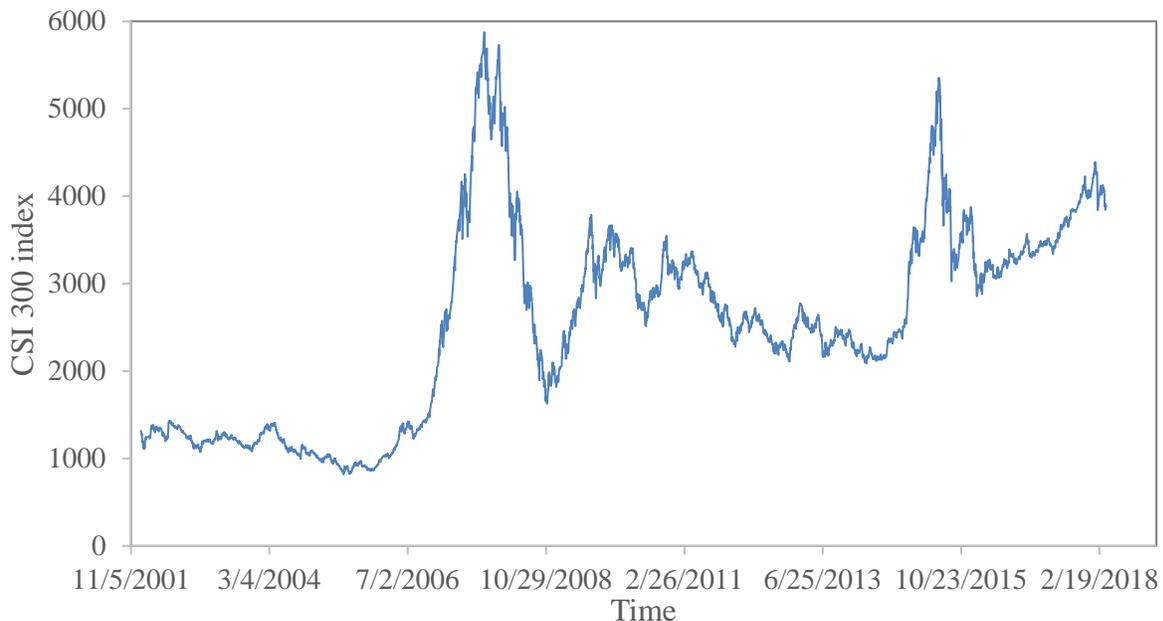

Figure 1. Evolution of the price trajectories of the CSI 300 index over the interval of this analysis

## 2. Methodology

*2.1 The Log-Periodic Power Law Singularity (LPPLS) Model*

The LPPLS Model, originally called as the Johansen-Leoit-Sornette (JLS) model, was initially proposed by Johansen, Ledoit and Sornette [4]. In this section, the derivation of the LPPLS model is recalled based on the original work [4]. The LPPLS model stems from a risk neutral rational agent with rational expectations and ignoring the arbitrage, dividends, the interest rate, risk aversion, information asymmetry and the market clearing condition. The rise of the expected asset price must compensate for the expected risk, implying the asset price follows a martingale process, i.e. $E_t[p(t')] = p(t), \forall t' > t$, where $p(t)$ denotes the asset price at the time $t$ and $E_t[\cdot]$ represents the conditional expectation given all previous data before and up to the time $t$. The occurrence of a crash or change can be modelled as a discontinuous jump process $j$ with the value of 0 before the crash and 1 after the crash occurrence at the critical time $t_c$. Due to the random nature of the crash occurrence, the $t_c$ can be modeled by the cumulative distribution function $Q(t)$, the probability density function $q(t) = dQ/dt$, and a crash hazard rate $h(t) = q(t)/[1 - Q(t)]$, which is the probability per unit of time of the crash taking place in the next instant conditional on the fact that it has not yet happened. Because $h(t)dt$ is the probability that the crash occurs between $t$ and $t + dt$ given the crash has not yet happened, the expectation of $dj$ can be determined as: $E_t[dj] = 1 \times h(t)dt + 0 \times (1 - h(t)dt) = h(t)dt$. For simplicity, it is assumed that the asset price falls during a crash at a fixed percentage $k \in (0,1)$. Then, the asset price dynamics before the crash occurring can be given by:

$$dp = \mu(t)p(t)dt - kp(t)dj + \sigma(t)p(t)dW \Rightarrow$$

$$E_t[dp] = \mu(t)p(t)dt - kp(t)E_t(dj) + \sigma(t)p(t)E(dW) = \mu(t)p(t)dt - kp(t)h(t)dt \quad (1)$$

where $\mu(t)$ is the time-dependent return, $\sigma(t)$ is the volatility, $dW$ is the infinitesimal increment of a standard Wiener process with zero mean and variance equal to $dt$. Under the assumption of no arbitrage and rational expectations, the conditional expectation of the price dynamics $E_t[dp]$ is zero as the price process satisfies the martingale condition, and so that $\mu(t)p(t)dt - kp(t)h(t)dt = 0$, yielding $\mu(t) = kh(t)$ implying the return $\mu(t)$ is proportional to the risk of crash quantified by its crash hazard rate $h(t)$. Due to the existence of noise traders with herding behavior, $h(t)$ plays a role of the driver of the bubble growing progressively on the no-arbitrage condition, leading to an instantaneous return $\mu(t)$ that grows together with $h(t)$ in order to remunerate investors who are willing to invest in a risk asset [11]. Substituting the equality of the return $\mu(t)$ into Equation (1), the asset price dynamics, conditioned on the fact that no crash occurs, can be simplified as:

$$dp = kh(t)p(t)dt - kp(t) \times 0 + \sigma(t)p(t)dW = kh(t)p(t)dt + \sigma(t)p(t)dW \Rightarrow$$

$$\frac{dp}{p(t)} = kh(t)dt + \sigma(t)dW \quad (2)$$

Its conditional expectation leads to $E_t[dp/p(t)] = kh(t)dt$ with the solution as follows:

$$E_t\left[\ln\left[\frac{p(t)}{P(t_0)}\right]\right] = k\int_{t_0}^{t} h(t')dt' \quad (3)$$

To model the behavior the asset price before a crash, it is necessary to specify the key variable: the crash hazard rate $h(t)$, which quantifies the probability that a large number of agents will assume the same sell position simultaneously resulting in the imbalance of financial market unless the asset price decrease substantially. In order to capture the imitative local micro-interactions, Johansen, Ledoit and Sornette [4] proposed a model in which each agent $i$ can have only two possible states $s_i$: "buy" ($s_i = +1$) or "sell" ($s_i = -1$). The state of agent $i$ at a given point in time is given by the following Markov process:

$$s_i = \text{sign}\left(K \sum_{k \in N(i)} s_j + \sigma \varepsilon_i\right) \quad (4)$$

where $\text{sign}(\cdot)$ represents the sign function with the value of +1 (-1) for positive (negative) numbers, $K$ is a positive constant qualifying the coupling strength between agents, $N(i)$ is the number of agents who influences agent $i$, $s_j$ is the current state of agent $s_j$, $\sigma$ is the tendency toward idiosyncratic behavior for all agents, $\varepsilon_i$ is the random draw from a standard normal distribution. The order $K/\sigma$ determines the outcome if order in the network wins. When order wins, the agents will imitate their close neighbors, resulting in the spreading of imitation in the whole network, and eventually causing a crash. When a crash occurs, $K$ will approach the critical value $K_c$, and all the agents will have the same state, either +1 or -1.

As Blanchard [24] pointed out, the higher the probability of a crash, the faster the price before the occurrence of crash should grow to satisfy the martingale condition, so that the investor induced to hold an asset with increasing risk of crash should be compensated by the chance of higher return. At this point, Johansen, Ledoit and Sornette [4] assumes that the behavior of the variable close to a critical point can be described by a power law, and the susceptibility of the critical system qualifying the degree of sensitivity of a system subjected to an external perturbation is expressed as $\chi \approx A(K_c - K)^{-\gamma}$ where $A$ is a positive constant (=7/4 for the bi-dimensional Ising model) and $\gamma$ is the positive critical exponent of the susceptibility. The susceptibility $\chi$ describes the chance that a large group of agents suddenly reach an agreement given the existent external influence in the network. In the 2-D Ising model, the interconnection of investors is only considered in a uniform way. However, in real modern financial market constituted of an ensemble of the investors which substantially differs in size ranging from individuals to gigantic professional funds, the interacting investors are organized inside a hierarchical network, where they locally influence each other at different levels. In order to appropriately represent the current structure of financial market, Johansen, Ledoit and Sornette [4] proposed a Hierarchical Diamond Lattice (HDL) to model the rational imitation of the investors. The structure of HDL is created by starting with a pair of linked traders and then substituting each link with a new diamond with four links and two new nodes diagonally opposite each other. This operation is repeated until the stopping criterion is satisfied. After $n$ iterations, there will be $\frac{2}{3}(2 + 4^n)$ traders and $4^n$ links among them. The HDL has the similar basic properties with the rational imitation model based on the bi-dimensional network. The only crucial difference is the that the critical exponent of the susceptibility $\gamma$ can be a complex number in HDL. A version of HDL was solved by Derrida, De Seze and Itzykson [25] and the general solution is given by:

$$\chi \approx \text{Re}[A_0(K_c - K)^{-\gamma} + A_1(K_c - K)^{-\gamma+i\omega} + \cdots]$$
$$\approx A'_0(K_c - K)^{-\gamma} + A'_1(K_c - K)^{-\gamma} \cos[\omega \ln(K_c - K) + \phi] + \cdots \quad (5)$$

where $A'_0, A'_1, \omega$ and $\phi$ are real numbers, and $\text{Re}[\cdot]$ denotes the real part of a complex number. The oscillations correct the pure pow law singularity, accounting for the underlying approximate discrete scale invariance of the financial price dynamics [26]. The oscillations are called "log-periodic" because they are periodic in logarithm of the variable $(K_c - K)$ and the angular log-frequency is $\frac{\omega}{2\pi}$. When the oscillations reach the critical time, their frequency explodes, leading to the accelerating oscillations. Accounting for this mechanism, the crash hazard rate is assumed to behave in a similar way to the susceptibility in the neighborhood of the critical point. Therefore, the hazard rate has the following form:

$$h(t) \approx \alpha(t_c - t)^{m-1}(1 + \beta \cos[\omega \ln(t_c - t) + \phi]) \quad (6)$$

where $\alpha, \beta, \phi, m, \omega$ and $t_c$ are parameters. This expression of the hazard rate shows that the risk of a crash per unit of time increase drastically when the interaction among investors increase before the occurrence of crash. Substituting the hazard rate in Equation (6) into the solution of the conditional expectation of the asset price in Equation (3), we get the evolution for the asset price before a crash, which is known as the LPPLS formula:

$$\text{LPPLS}(t) \equiv E_t[\ln p(t)] = A + B(t_c - t)^m\{1 + C\cos[\omega \ln(t_c - t) + \phi]\} \quad (7)$$

where $A > 0$ is the expected value of the $\ln p(t_c)$ at the critical time $t_c$, $B = -k\alpha/m < 0$ for a positive bubble is the decrease in $\ln p(t)$ over the time unit if $C$ is close to zero before a crash, $C = -k\alpha\beta/\sqrt{m^2 + \omega^2}$ is the proportional magnitude of the oscillations around the power law singular growth, $0 < m < 1$ is the exponent of the power law growth, $\omega$ is the angular log-frequency of the oscillation during a bubble, and $0 < \phi < 2\pi$ is a phase parameter. The Equation (7) is the fundamental equation of LPPLS formula describing the evolution of asset prices before a crash occurs and it has been proposed in different forms in several papers, e.g., Sornette [27] and Lin, Ren and Sornette [17].

Two common remarkable characteristics of the most speculative bubble are well documented in both developed and emerging stock markets, i.e., (1) a faster-than-exponential (or super-exponential) growth of the stock market, which ends when the bubble regime changes and (2) accelerating oscillations when approaching to the critical time of the bubble [3, 5, 28]. Both the significant features can be well captured by the LPPLS model in Equation (7). The feature of super-exponential growth of the bubble can be described by the power law singular component $A + B(t_c - t)^m$, which embodies the positive feedback mechanism of a bubble development. To ensure the super-exponential growth, it is required that $0 < m < 1$. The condition $m > 0$ makes sure that the price remains finite at the critical time $t_c$, while $m < 1$ expresses that a singularity exists. The positive bubble when the price of asset is arising is characterized by $B < 0$, while the negative bubble when the price is falling is featured by $B > 0$. $A > 0$ ensures the price of asset is positive. The asset price dynamics of anti-bubble can be obtained by replacing $t_c - t$ by $t - t_c$. The characteristic of accelerating oscillations of the bubble is captured by the component $C(t_c - t)^m \cos[\omega \ln(t_c - t) + \phi]$, which represents the tension and competition between the value investors and the noise traders resulting in the deviation of the market price around the super-exponential growth in the form of oscillations that are periodic in the logarithm of the time to $t_c$. The term $C(t_c - t)^m$ describes the fact that the amplitude of the accelerating oscillation is falling to zero at the critical time $t_c$. The term $\omega \ln(t_c - t)$ represents the local frequency of the log-periodic oscillations is accelerating to infinite at the critical time $t_c$. The parameter $\phi$ is related to the characteristic time unites for the oscillations. It should be noted that the critical time $t_c$ is the most probable time for a change in regime at which the growth rate of the asset price changes. The regime change is often but not necessarily the time of a bubble crash. A change in regime refers to a change from super-exponential growth to an exponential or lower growth with the end of the accelerating oscillations.

*2.2 LPPLS calibration*

The original LPPLS formula in Equation (7) consists of 3 linear parameters $(A, B, C)$ and 4 nonlinear parameters ($t_c, m, \omega, \phi$). A common method of calibration for the LPPLS model is the ordinary least squares method. The 3 linear parameters $(A, B, C)$ are enslaved in the fitting algorithm to simplify the calibration and then estimated from the solved solutions of the 4 nonlinear parameters ($t_c, m, \omega, \phi$). However, the calibration of the LPPLS model by minimizing the nonlinear multivariate least squares functions is a non-trivial task because of the relatively large number of parameters and the strong nonlinear structure of the model and the multiple local minima rendering the local optimization algorithms getting trapped. The solution for the global minimum may not be attainable by utilizing metaheuristic methods such as taboo search (Cvijovic & Klinowski, 1995) or genetic algorithm [29]. In order to reduce the number of nonlinear parameters and lessen the interdependence between the angular log-frequency $\omega$ and the phase $\phi$, Filimonov and Sornette [15] proposed transforming the LPPLS formula to reduce the number of nonlinear parameters from 4 to 3 at the cost of increasing the number of linear parameters from 3 to 4 as the following:

$$\text{LPPLS}(t) \equiv E_t[\ln p(t)] = A + B(t_c - t)^m + C_1(t_c - t)^m \cos[\omega \ln(t_c - t)] \\ + C_2(t_c - t)^m \sin[\omega \ln(t_c - t)] \quad (8)$$

Here $C_1 = C\cos\phi$ and $C_2 = C\sin\phi$. The phase $\phi$ is contained by $C_1$ and $C_2$. The cost function in the least-squares method can be described as:

$$F(t_c, m, \omega, A, B, C_1, C_2) = \sum_{i=1}^{N} [\ln p(\tau_i) - A - B(t_c - \tau_i)^m - C_1(t_c - \tau_i)^m \cos(\omega \ln(t_c - \tau_i)) \\ - C_2(t_c - \tau_i)^m \sin(\omega \ln(t_c - \tau_i))]^2 \quad (9)$$

where $\tau_1 = t_1$ and $\tau_N = t_2$.

Subordinating the 4 linear parameters $A, B, C_1$ and $C_2$ to the 3 nonlinear parameters $t_c, m, \omega$, the nonlinear optimization problem is: $\{\hat{t}_c, \hat{m}, \hat{\omega}\} = arg \min_{t_c, m, \omega} F_1(t_c, m, \omega)$. Here $F_1(t_c, m, \omega) = \min_{A, B, C_1, C_2} F_1(t_c, m, \omega, A, B, C_1, C_2)$. The linear parameters can be solved by:

$$\begin{pmatrix} N & \sum f_i & \sum g_i & \sum h_i \\ \sum f_i & \sum f_i^2 & \sum f_i g_i & \sum f_i h_i \\ \sum g_i & \sum f_i g_i & \sum g_i^2 & \sum h_i g_i \\ \sum h_i & \sum f_i h_i & \sum g_i h_i & \sum h_i^2 \end{pmatrix} \begin{pmatrix} \hat{A} \\ \hat{B} \\ \hat{C}_1 \\ \hat{C}_2 \end{pmatrix} = \begin{pmatrix} \sum \ln p_i \\ \sum f_i \ln p_i \\ \sum g_i \ln p_i \\ \sum h_i \ln p_i \end{pmatrix} \quad (10)$$

where $f_i = (t_c - \tau_i)^m$, $g_i = (t_c - \tau_i)^m \cos(\omega \ln(t_c - \tau_i))$, and $h_i = (t_c - \tau_i)^m \sin(\omega \ln(t_c - \tau_i))$. The cost function of the transforming LPPLS model is characterized by good smooth properties, leading to the dramatic reduction of the complexity and tremendous improvement of stability in the fitting procedure, so that the metaheuristic methods are no longer necessary, and the fitting efficiency significantly increases. In this study, the covariance matrix adaptation evolution strategy (CMA-ES) is adopted to search the best estimation of the three nonlinear parameters ($t_c, m, \omega$) by minimizing the residuals (the sum of the squares of the differences) between the fitted LPPLS model and the observed price time series. The CMA-ES proposed by [30] ranks among the most successful evolutionary algorithms for real-valued single-objective optimization and is typically applied to difficult nonlinear non-convex black-box optimization problems in continuous domain and search space dimensions between three and a hundred. Parallel computing is applied to expedite the fitting process with remarkable reduction in computation time.

*2.3 LPPLS confidence indicator*

The LPPLS confidence indicator was introduced by Sornette, Demos, Zhang, Cauwels, Filimonov and Zhang [11] and is also one of key indicators in Financial Crisis Observatory (FCO) at ETH Zurich. The LPPLS confidence indicator is defined as the fraction of fitting windows in which the LPPLS calibrations satisfy the specified filter conditions. It is used to measure sensitivity of observed bubble pattern to the time interval between the end time and the start time in the fitting windows ($dt = t_2 - t_1$). A large value of the LPPLS confidence indicator indicates a more reliable LPPLS pattern. A small value of the indicator signals a possible fragility since the LPPLS pattern is presented in a few fitting windows.

The LPPLS confidence indicator for a specified data point $t_2$ (corresponding to a fictitious "present") can be obtained by the following five steps: (1) create the fitting time windows by shrinking in terms of $t_1$ moving toward the fixed endpoint $t_2$ with a step of $dt_1$, (2) determine the search space in the calibration procedure, (3) calibrate the LPPLS model for each fitting time window, (4) specify the filter conditions and summarize the number of fitting windows in which satisfy the specified filter condition, and (5) calculate the LPPLS confidence indicator from dividing the number of time windows satisfying the specified filter condition by the total number of the fitting windows.

In this study, the length of the shrinking time windows $dt = t_2 - t_1$ is adopted to decrease from 750 trading days to 50 trading days in steps of 5 trading days. Thus, 141 fitting windows are obtained for each $t_2$. In order to minimize fitting problems and address the sloppiness of the model, we adopt the following search space:

$$m \in [0,1], \omega \in [1, 50], t_c \in \left[t_2, t_2 + \frac{t_2 - t_1}{3}\right], \frac{m|B|}{\omega\sqrt{C_1^2 + C_2^2}} \geq 1 \qquad (11)$$

The condition $t_c \in [t_2, t_2 + (t_2 - t_1)/3]$ ensures that the predicted critical time $t_c$ should be after the endpoint $t_2$, and should not be too far away from $t_2$ since the predictive capacity degrades far beyond $t_2$ [10]. The Damping parameter satisfies $m|B|/\left(\omega\sqrt{C_1^2+C_2^2}\right) \geq 1$ under the condition that the crash hazard rate $h(t)$ is non-negative by definition [31]. After calibrating the LPPLS models, the solutions should be filtered by the stricter conditions:

$$m \in [0.01, 0.99], \omega \in [2, 25], t_c \in \left[t_2, t_2 + \frac{t_2 - t_1}{5}\right], \frac{\omega}{2}\ln\left(\frac{t_c - t_1}{t_c - t_2}\right) \geq 2.5,$$

$$\max\left(\frac{|\hat{p}_t - p_t|}{p_t}\right) \leq 0.15, \ p_{lomb} \leq \alpha_{sign}, \ \ln(\hat{p}_t) - \ln(p_t) \sim AR(1) \quad (12)$$

The filter conditions are derived from the empirical evidence gathered in investigations of previous bubbles [10, 11] and are the stylized features of LPPLS model. The condition for the number of oscillations (half-periods) of the log-periodic component $(\omega/\pi)\ln[(t_c - t_1)/(t_c - t_2)] \geq 2.5$ is implemented to distinguish a genuine log-periodic signal from one that could be generated by noise [32]. The condition of the maximum relative error $\max(|\hat{p}_t - p_t|/p_t) \leq 0.15$ ensure the fitted price of an asset $\hat{p}_t$ should be not too far from the actual asset price $p_t$. The condition $P_{lomb} \leq \alpha_{sig}$ ensures the logarithm-periodic oscillations in the fitting the logarithm of prices to the LPPLS model by applying the Lomb spectral analysis for the series of detrended residual $r(t) = (t_c - t)^{-m}(\ln[p(t)] - A - B(t_c - t)^m)$ [33]. The probabilities that the maximum peak occurred by chance $P_{lomb}$ is less than the specified significant level $\alpha_{sig}$, indicating the existence the logarithm-periodic oscillations in the fitting LPPLS model. The $\ln(\hat{p}_t) - \ln(p_t) \sim AR(1)$ condition ensures that the LPPLS fitting residuals can be modeled by a mean-reversal Ornstein-Uhlenbeck (O-U) process when the logarithmic price in the bubble regime is attributed to a deterministic LPPLS component [17]. Since the test for the O-U property of LPPLS fitting residuals can be translated into an AR(1) test for the corresponding residuals, both the Phillips-Perron unit-root test and Dickey-Fuller unit-root test are used to check the O-U property of LPPLS fitting residuals. The 10% significant level is adopted for the tests in this study. Only the calibrations satisfying filter conditions given in Equation (12) are considered valid and the rest are discarded.

## 3. Empirical analysis

In the following two subsections, we present the detection analysis of the Chinese Stock Market bubbles using the LPPLS confidence indicator as described in Section 2, followed by the post-mortem analysis of these bubbles.

### 3.1 LPPLS bubble identification

In this study, we have collected the daily data on the Shanghai Shenzhen CSI 300 stock market index from January 4, 2002 through April 2, 2018 for a total of 3,939 observations. These data come from the Bloomberg Financial Database. We adopted the length of the shrinking time windows $t_2 - t_1$ decreasing from 750 trading days to 50 trading days in steps of 5 trading days and the endpoint $t_2$ moving from March 1, 2005 through April 2, 2018 in steps of 5 trading days to generate 638 $t_2$. Since there are 141 fitting windows for each $t_2$, a total of 89,958 fitting windows are generated in this study. The value of the LPPLS confidence indicator at a given time $t_2$ is causal because it is estimated based only on data prior to that time. The LPPLS confidence indicators for a series of varying $t_2$ provide useful insights into the time development of the bubble signal.

Both the positive and negative bubbles in the Chinese stock market are detected in this study. The positive bubbles are associated with the upwardly accelerating price increases, and susceptible to regime changes in the form of crashes or volatile sideway plateaus, while the negative bubbles are associated with the downwardly accelerating price decreases, and are susceptible to regime changes in the form of rallies or volatile sideway plateaus. Figures 2-4 show the LPPLS confidence indicator for positive bubbles in red along with the CSI 300 index in blue from 3/1/2005 to 4/2/2018. Figures 5–7 presents the LPPLS confidence indicator for negative bubbles in red along with the CSI 300 index in blue from 3/1/2005 to 4/2/2018. These figures indicate the confidence level of the observed LPPLS bubble patterns. The LPPLS confidence indicator marks bubble by measuring the sensitivity of the bubble pattern to the selected starting time. When the LPPLS bubble pattern exists in more time windows for a given "present" time, the LPPLS confidence indicator has a higher value. The value of LPPLS confidence indicator can be up to one if the bubble pattern exists in most of the analyzed time windows and presents almost no sensitivity to the choice of the time windows. When the bubble pattern is only observed in a few of time windows, the LPPLS confidence indicator may have a value close to zero which indicates the over-fitting risk and needs careful consideration for the results.

As shown in Figure 1, a cluster of bubble patterns are detected from January 2007 to October 2007, indicating a positive bubble may have been born and developed over time since the value of LPPLS confidence indicator has a dramatic increase. Multiples peaks of the confidence indicator with large value are observed from May to October 2007 and the highest value of indicator is up to 0.27, representing the observed bubble signals are reliable and the regime change may occur in form of crash or volatile sideway plateaus, so that the growth rate of CSI 300 index would be changed from super-exponential growth to an exponential or lower growth. The diagnostic of the presence of possible bubble is confirmed by the Chinese stock bubble of 2007, in which the CSI 300 index reaches the historical peak of 5877.2 on October 16, 2007. The Chinese stock bubble of 2007 corresponded to an approximate 314.3% growth in just one year. After the bubble crashed, the CSI 300 index lost more than 70% from the historical high during the period from October 2007 to October 2008.

In Figure 2, a cluster of positive bubble signals are diagnosed from March to August in 2009 and the LPPLS confidence indicator reaches the peak of 0.10 on July 28, 2009. During the 2009 Chinese stock bubble, the CSI 300 index has risen more than 130% from November 4, 2008 to August 3, 2009. Following the bubble crash, the index fell by over 25% in August 31, 2009. Thus, the bubble patterns indeed detect the development and crash of the 2009 Chinese stock bubble.

From Figure 3, two clusters of positive bubble signals can be seen from November 7, 2014 to January 13, 2015, and from April 29, 2015 to June 11, 2015, separately. On June 11, 2015, the LPPLS confidence indicator reaches the top of 0.118, indicating the high risk of regime change. The first diagnostic of a "bubbly" CSI 300 index occurred November 2014 and persisted until January 2015, when a change of regime indeed occurred. Afterwards, the bubble pattern reappears and becomes stronger on April 2015 and persisted until the eventual burst of the 2015 Chinese Stock Market bubble. The CSI 300 index has suffered more than 42% drop from the peak on June 12, 2015 to the bottom on August 26, 2015. It is noted that some bubble signals appear from October to November in 2016 and September in 2017, implying the potential bubble may be emerging and a significant change of regime may occur in the future.

Figure 4 shows two clusters of negative bubbles. The first one is from March 22, 2005 to August 2, 2005 with the peak of 0.085 occurs at July 12, 2005. This cluster captures the Chinese stock market negative bubbles in 2005. The CSI 300 index starts to fall from 1,410.43 on April 9, 2004 to the historical lowest value 824.1 on July 11, 2005. The second cluster starts on April 8, 2008, and end on November 10, 2008 with the confidence indicator value 0.064. The CSI 300 index falls 4,104 points from January 14, 2008 to November 4, 2008 (71.6% decline).

There are two main clusters of negative bubbles in Figure 5. The first cluster starts on September 14, 2011 and culminates on January 13, 2012 with the start of the rebound. In the negative bubble, the CSI 300 index has suffered more than 32% drop from April 15, 2011 to January 5, 2012. The second cluster is from July 30, 2012 to December 10, 2012. The LPPLS confidence indicator surges to 0.156 on December 10, 2012, followed by the rebound of the price. The CSI 300 index fell from 2,717.8 on May 7, 2012 to 2108.9 on December 3, 2012 and then rebound to 2,673.3 on February 28, 2013.

Figure 6 presents two clusters of negative bubbles from March 28, 2014 to July 17, 2014 and from August 21, 2015 to September 8, 2015. As shown in Figure 1, the CSI 300 index has a valley from March to July 2014, and then rises dramatically in the next one year to form the well-known 2015 Chinese Stock Market bubble. The 2015 Chinese Stock Market bubble dropped to the bottom on September 8, 2015 and then the regime changes in form of rebound.

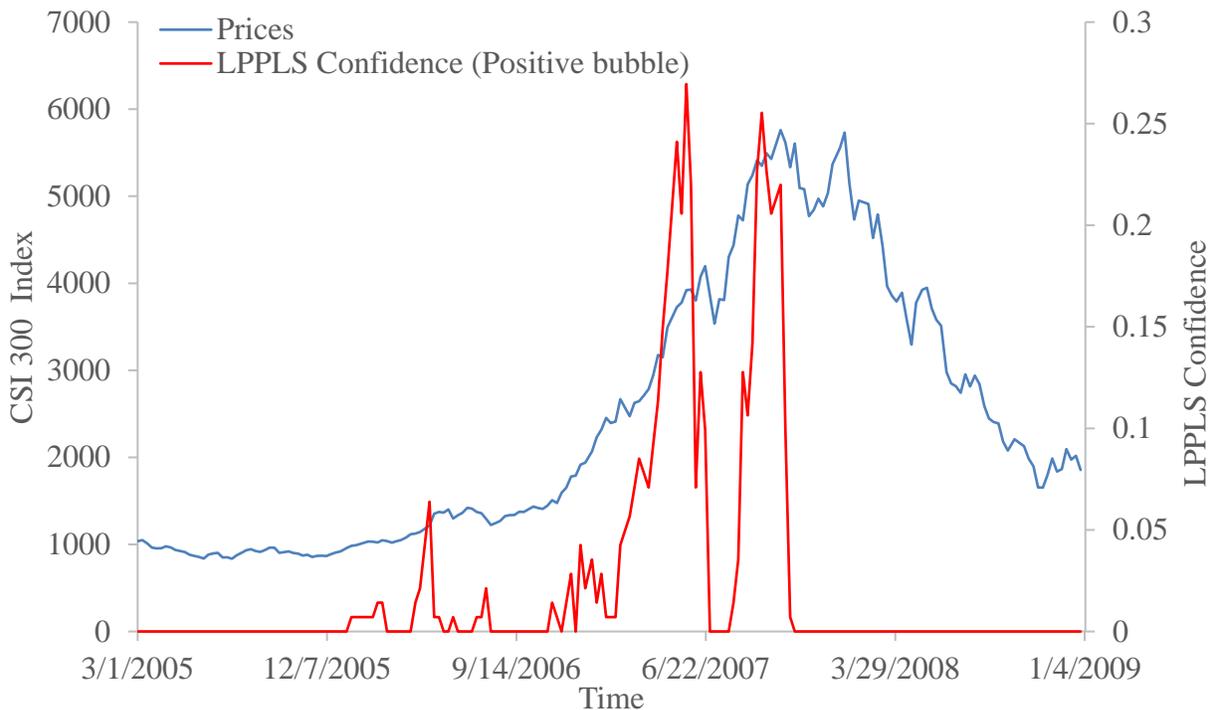

Figure 1. LPPLS confidence indicator for positive bubbles in red (right scale) together with the CSI 300 index in blue (left scale) from 3/1/2005 to 12/29/2008

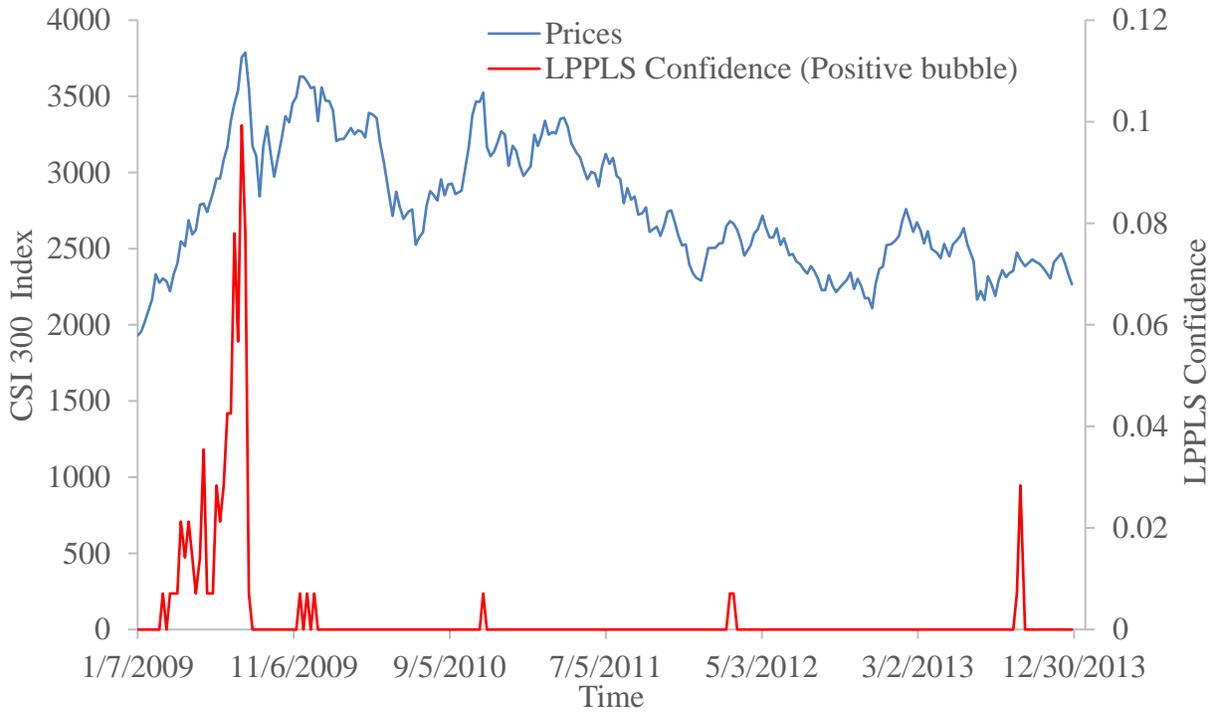

Figure 2. LPPLS confidence indicator for positive bubbles in red (right scale) together with the CSI 300 index in blue (left scale) from 1/7/2009 to 12/26/2013

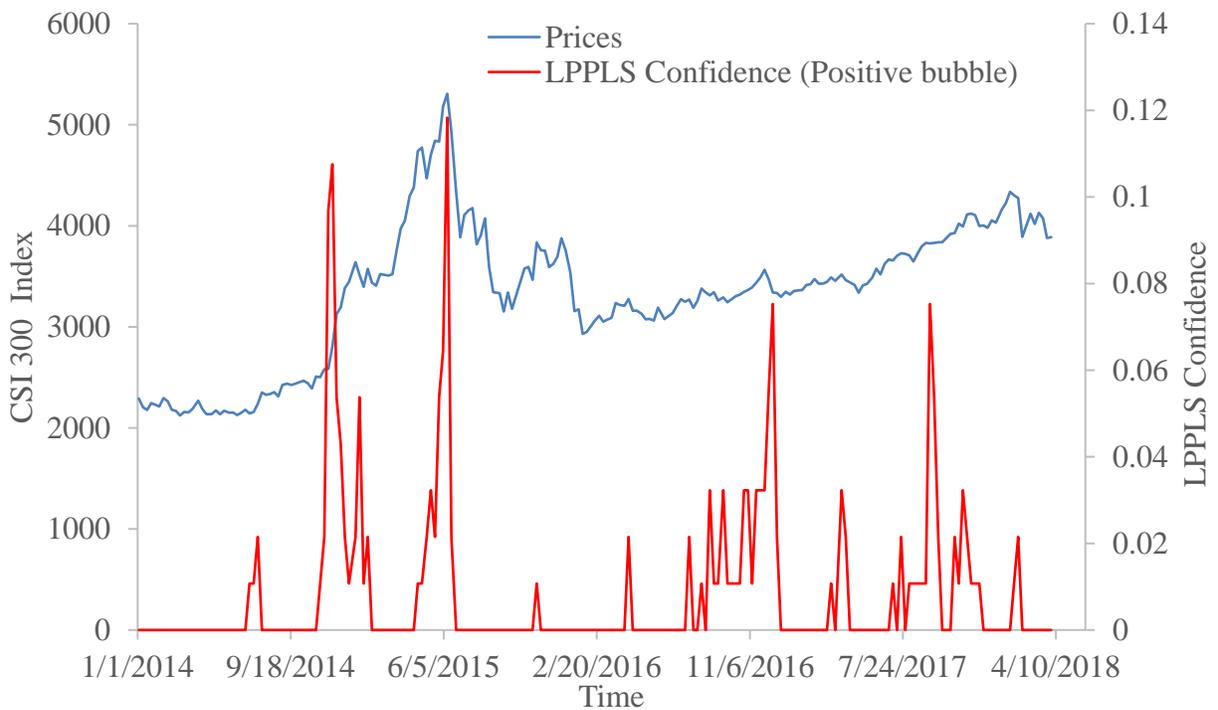

Figure 3. LPPLS confidence indicator for positive bubbles in red (right scale) together with the CSI 300 index in blue (left scale) from 1/3/2014 to 4/2/2018

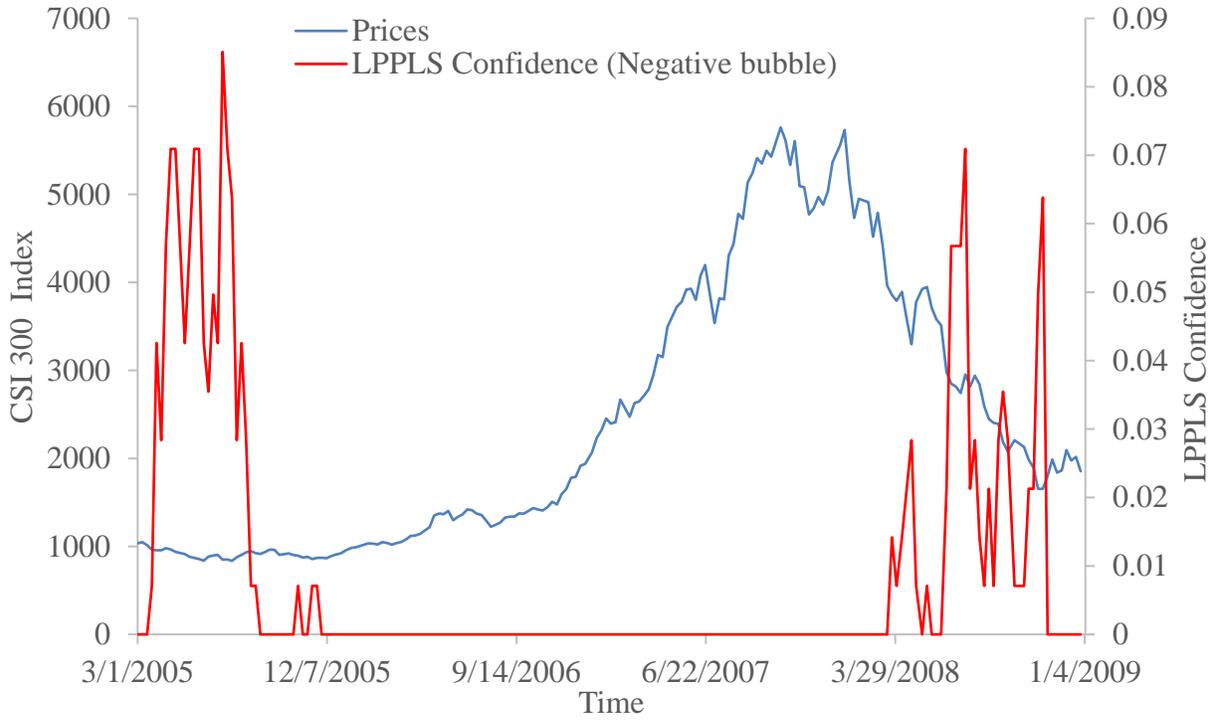

Figure 4. LPPLS confidence indicator for negative bubbles in red (right scale) together with the CSI 300 index in blue (left scale) from 3/1/2005 to 12/29/2008

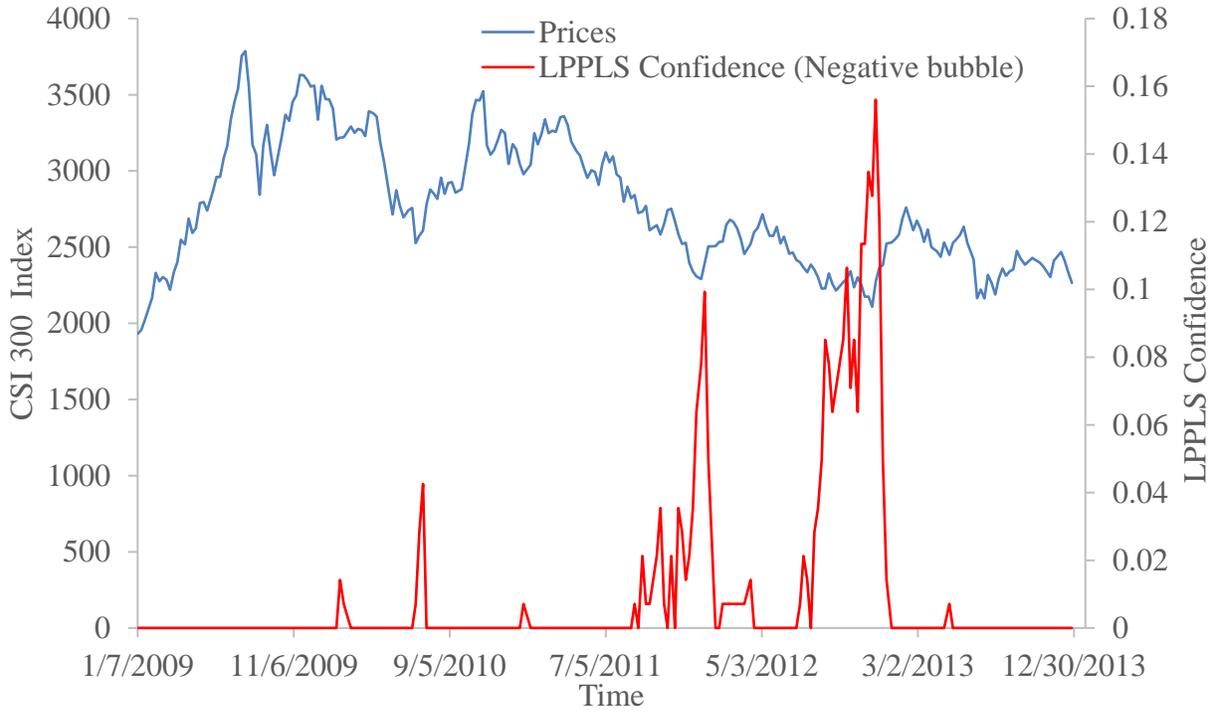

Figure 5. LPPLS confidence indicator for negative bubbles in red (right scale) together with the CSI 300 index in blue (left scale) from 1/7/2009 to 12/26/2013

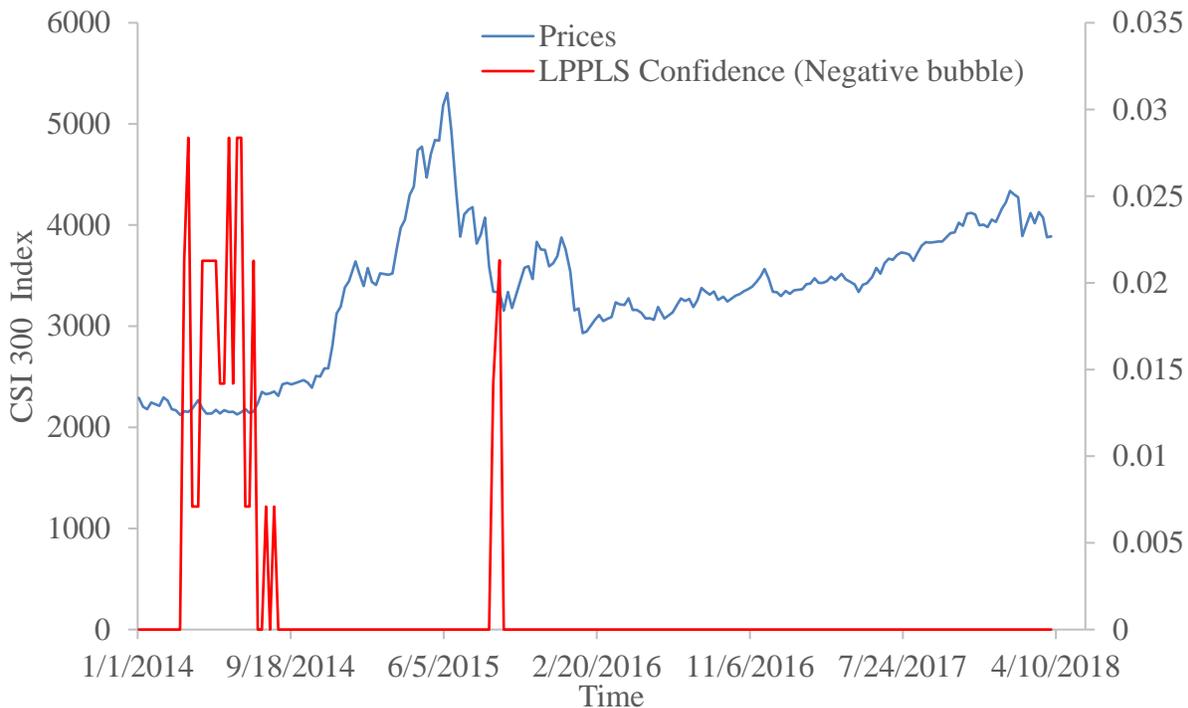

Figure 6. LPPLS confidence indicator for negative bubbles in red (right scale) together with the CSI 300 index in blue (left scale) from 1/3/2014 to 4/2/2018

*3.2 Post-mortem analysis for the bubbles*

This section presents the additional results about the two "well-known" Chinese stock market bubbles: the 2007 and the 2015 bubbles. This analysis provides more detailed information of the LPPLS methodology for detecting the bubbles and their termination.

Figure 7 shows the probability density distribution of the predicted $t_c$'s as well as the estimated beginning $t_1$'s for the Chinese stock bubble of 2007. This is obtained by scanning over 141 $t_1$'s from a maximum 750 trading days to a minimum of 50 trading days in steps of 5 trading days prior to each the end $t_2$ in which the fitting windows passed the filter conditions in Equation (12) are collected, and repeating this procedure for different $t_2$ in steps of 5 trading days, and eventually generating the probability density distribution by statistical analysis on the chosen fitting windows. The ranges of $t_1$ and $t_2$ for the 220 selected fitting windows are from August 30, 2005 to June 28, 2007 and from August 23, 2007 to October 11, 2007, respectively. The optimal values for the bubble starting date $t_1$ are represented by the probability density distribution $pdf(t_1)$ in green. It can be seen that the $pdf(t_1)$ is concentrated in the time interval where the CSI 300 index starts to super-exponentially accelerate. This allows us to determine the beginning of the Chinese stock bubble of 2007 as early as August 30, 2005. The forecasted critical time $t_c$ depicted by the probability density distribution $pdf(t_c)$ in red presents a strong probability measure at the time of the crash.

As shown in Figure 7, the 20%/80% and 5%/95% quantile range of values of the crash dates $t_c$ for the Chinese stock bubble of 2007 are from September 21, 2007 to October 22, 2007, and from August 30, 2007 to November 12, 2007. The observed market peak date for the CSI 300 index is October 16, 2007, which lies in the quantile ranges of the predicted crash dates $t_c$ fitted based on data before the actual stock market crash. Figure 8 also presents three typical fitting examples corresponding to $t_1$= 18 October 2005 and $t_2$= 13 September 2007, $t_1$= 20 November 2006 and $t_2$= 11 October 2007, and $t_1$= 19 April 2007 and $t_2$= 20 September 2007, which represent the different time scale windows, respectively.

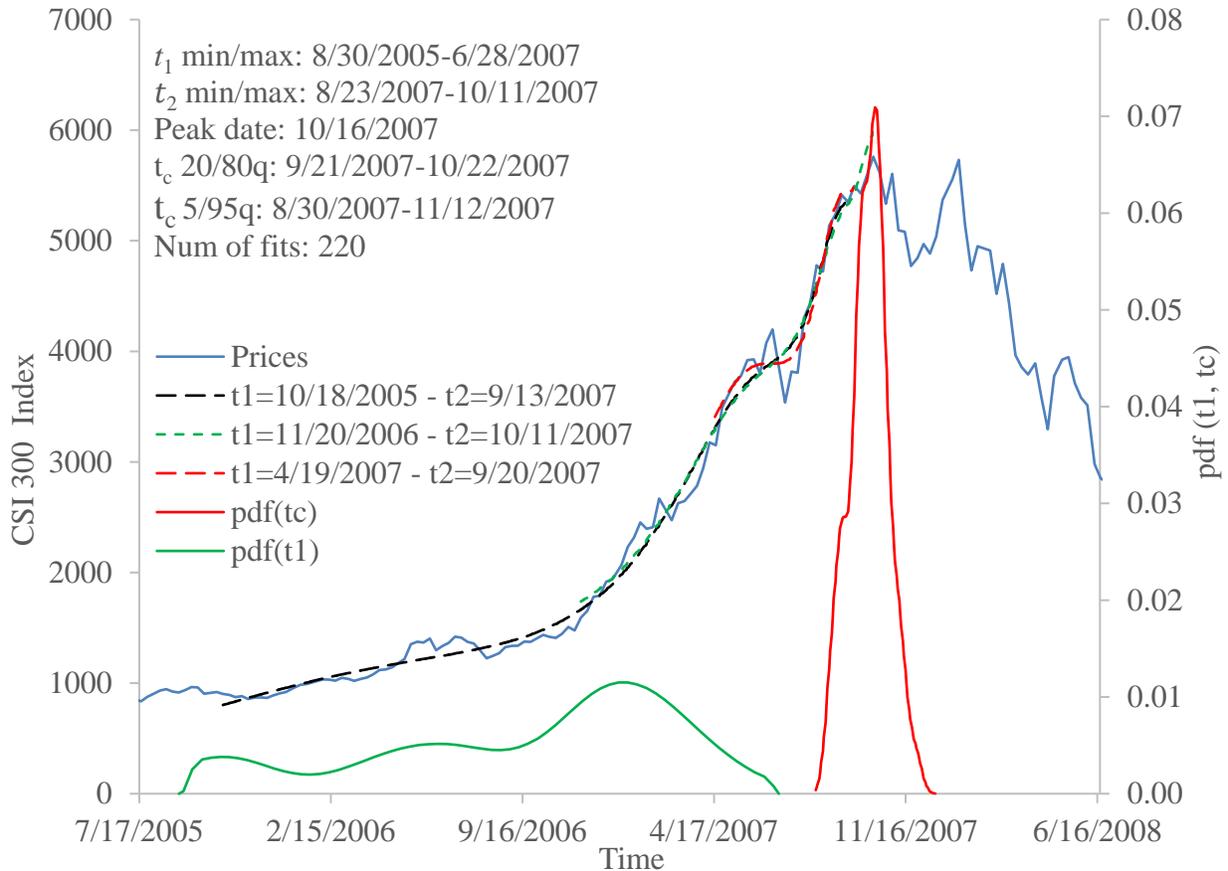

Figure 7. The probability density distributions $pdf(t_1, t_c)$ for the Chinese stock bubble of 2007 (right scale) together with the CSI 300 index in blue (left scale) from 7/17/2005 to 6/16/2008

The probability density distribution of the predicted $t_c$'s as well as the estimated beginning $t_1$'s for the 2015 Chinese Stock Market bubble is shown in Figure 8. The ranges of $t_1$ and $t_2$ for the 48 selected fitting windows are from February 7, 2015 to March 17, 2015 and from April 22, 2015 to June 11, 2015, respectively. The probability density distribution $pdf(t_1)$ in green presents the optimal values for the bubble starting date $t_1$. It is observed that the $pdf(t_1)$ is skew negatively and the mass of the distribution is concentrated on the right of the figure where the CSI 300 index has a super-exponentially growth. From Figure 8, we can see that the 2015 Chinese Stock Market bubble began as early as February 7, 2015. The probability density distribution $pdf(t_c)$ in red represents the predicted critical time $t_c$.

Figure 8 shows that the 20%/80% and 5%/95% quantile range of values of the crash dates $t_c$ for the 2015 Chinese Stock Market bubble are from June 10, 2015 to July 22, 2015, and from May 27, 2015 to August 6, 2015, respectively. The observed market peak date for the CSI 300 index (June 12, 2015) lies in the quantile ranges of the predicted crash dates $t_c$ fitted based on data before the actual stock market crash. Figure 8 also illustrates three typical fitting examples are corresponding to $t_1$= 14 April 2014 and $t_2$= 14 May 2015, $t_1$= 21 November 2014 and $t_2$= 14 May 2015, and $t_1$= 20 January 2015 and $t_2$= 4 June 2015 to represent the long, median and short time scale windows, respectively.

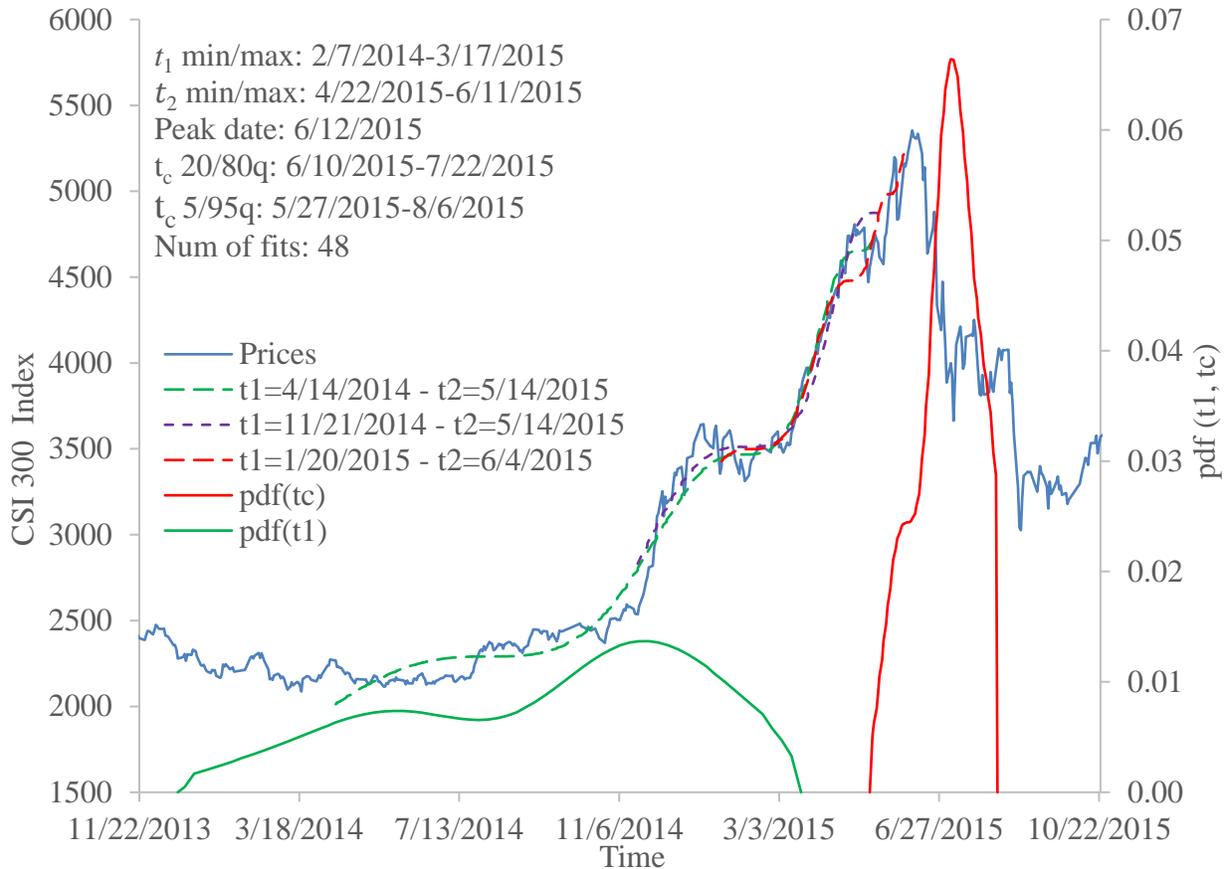

Figure 8. The probability density distributions $pdf(t_1, t_c)$ for the 2015 Chinese Stock Market bubble (right scale) together with the CSI 300 index in blue (left scale) from 11/22/2013 to 10/22/2015

## 4. Conclusions

In this study, we apply the LPPLS methodology to carry out the early causal identification of positive and negative bubbles in the Chinese stock market using the daily data on the Shanghai Shenzhen CSI 300 stock market index dated from January 2002 through April 2018. In order to improve the performance of the LPPLS confidence indicator, we adjust the search space to account for the damping condition of the LPPLS model and implement the stricter filter conditions for the qualification of the valid LPPLS fits by taking account of the maximum relative error, the Lomb log-periodic test of the detrended residual, and the unit-root tests of the logarithmic residual based on both the Phillips-Perron test and the Dickey-Fuller test. This study is the first of its kinds that identifies the existence of bubbles in the Chinese stock market using the daily data of CSI 300 index with the advance bubble detection methodology of LPPLS confidence indicator. Using only historical data, our analysis shows that the LPPLS detection strategy was able to forecast three periods of positive bubbles and four periods of negative bubbles occurred in the period from March 1, 2005 to April 2, 2018. The bubble periods detected by our methodology correspond to well-known historical events, implying the detection strategy based on the LPPLS confidence indicator has an outstanding performance in identifying the potential positive and negative bubbles in advance.

This study implements the post-mortem analysis for the two "well-known" Chinese stock market bubbles: the 2007 and the 2015 bubbles. The probability density distributions and quantile ranges of the predicted critical time $t_c$'s plus the estimated beginning time $t_1$'s provide a strong probability measure about the time of the crash and the beginning of the bubbles. It is observed that the probability density distribution of the estimated beginning time of bubbles appears to be skewed and the mass of the distribution is concentrated on the area where the price starts to have an obvious super-exponentially growth.

It can also be found that the regime shifts and changes are not a rare phenomenon and may occur more frequently in the future. This study shows that it is possible to detect the potential positive and negative bubbles and crashes ahead of time, which provides a prerequisite for limiting the bubble sizes and eventually minimizing the damages from the bubble crashes.